\documentclass[aps,pra,twocolumn,amsmath,amssymb,superscriptaddress,nofootinbib,longbibliography]{revtex4}
\usepackage{graphicx}
\usepackage{bm}
\usepackage{amssymb}
\usepackage{amsmath}
\usepackage{latexsym}
\usepackage{color}
\usepackage{siunitx}
\usepackage{graphicx} 
\usepackage{changes}
\definechangesauthor[name=yuri,color=red]{yuri}

\newcommand{\vc}[1]{\mbox{\boldmath $#1$}} 
\newcommand{\ind}[1]{_{#1}}    
\newcommand{\indrm}[1]{_{\mathrm {#1}}}    
\newcommand{\dirate}{{\mathcal D}}   
\newcommand{\dum}{{D_{\ind{t}}}}   
\newcommand{\sgn}{s}   
\newcommand{\dcomm}[1]{b_{\cup_{\ind{#1}}}}   

\newcommand{\bcomm}[1]{B_{\cup_{\ind{#1}}}}      
\newcommand{\gcomm}[1]{G_{\cup_{\ind{#1}}}}      
\newcommand{\fcomm}[1]{\dirate_{\cup_{\ind{#1}}}}
\newcommand{\yzplane}{diffraction}   
\definecolor{darkgreen}{rgb}{0,0.6,0}
\definecolor{darkred}{rgb}{0.8,0,0}
\definecolor{darkblue}{rgb}{0.,0,0.4}

\begin{document}  
\title{High-Luminosity  meV-Resolution Single-Shot Hard  X-ray Spectrograph for Cavity-Based X-ray Free-Electron Lasers}

\author{Keshab Kauchha}
\affiliation{Argonne National Laboratory, Lemont, Illinois, USA}

\author{Peifan Liu}
\affiliation{Argonne National Laboratory, Lemont, Illinois, USA}

\author{Paresh Pradhan}
\affiliation{Argonne National Laboratory, Lemont, Illinois, USA}

\author{Yuri Shvyd'ko} \email{shvydko@anl.gov}
\affiliation{Argonne National Laboratory, Lemont, Illinois, USA}

\begin{abstract}
Cavity-based x-ray free-electron lasers (CBXFELs) represent a possible
realization of fully coherent hard x-ray sources having high spectral
brilliance along with a narrow spectral bandwidth of $\simeq 1 - 50$~meV, a
high repetition pulse rate of $\simeq 1$~MHz, and good stability.  A
diagnostic tool is required to measure CBXFEL spectra with meV
resolution and high luminosity on a shot-to-shot basis.  We have designed a
high-luminosity single-shot hard x-ray spectrograph that
images 9.831-keV x-rays in a $\simeq 200$~meV spectral window with a
spectral resolution of a few meV.  The 
spectrograph is designed around angular dispersion of x-rays in Bragg diffraction  from
crystals.  It operates close to design specifications,
exhibiting a linear dispersion rate of $\simeq$~1.4~$\mu$m/meV and a $\simeq$~200-meV
window of high-fidelity spectral imaging. The experimentally demonstrated spectral resolution is $\simeq 20$~meV; this resolution is twice as low as expected from theory primarily because the spectrograph is highly sensitive to crystal angular instabilities. The experiment was performed at the bending magnet x-ray optics testing beamline 1-BM at the Advanced Photon Source.  
\end{abstract}

\maketitle


\section{Introduction and principles of spectrographs}
With the advent of high-gain, single-pass x-ray free-electron lasers
(XFELs) \cite{EAA10,SACLA,KMH17,EXFEL20} exhibiting extreme
brightness, transverse coherence, and ultra-short pulse length, a broad
range of new scientific applications became possible, extending from
investigation of femtosecond dynamics of atomic and molecular
systems~\cite{BBB13,CGI21} to detection of long-lived ultra-narrow nuclear
resonances \cite{SRK23}.  

Initially, XFELs were based on a
self-amplified spontaneous emission \cite{KS80,BPN84} (SASE) process
starting from shot noise and therefore having poor longitudinal
coherence of x-ray pulses. Over time, various approaches have been used to improve the longitudinal coherence. Among them have been various  self-seeding schemes
\cite{FSS97,SSSY,GKS11,HXRSS12,IOH19,NMO21,LGG23}. A somewhat different strategy is the cavity-based x-ray
free-electron laser (CBXFEL), which improves coherence by using x-ray feedback from a narrow-band
x-ray cavity, such as a low-gain x-ray FEL oscillator
(XFELO)~\cite{KSR08,KS09,LSKF11} or a high-gain x-ray regenerative
amplifier FEL (XRAFEL)~\cite{HR06,MDD17,FSS19,MHH20}. This strategy holds promise for producing high-brilliance x-rays not only  with full
coherence and a high repetition rate of $\simeq 1$~MHz but also with 
good stability.  The CBXFEL pulses are expected to have a narrow
energy bandwidth, which can be as small as a few meV for XFELOs,
although with about $10^2$ times smaller pulse energy of $\simeq 10~\mu$J
~\cite{KSR08,KS09,LSKF11}.

\begin{figure}[t!]
    \includegraphics[width=0.5\textwidth]{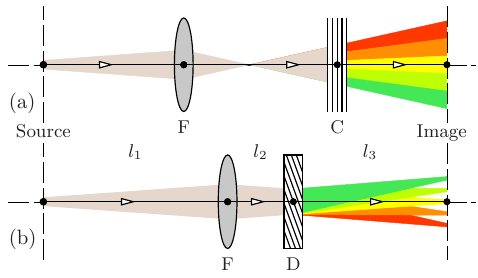}  
  \caption{Schematics of spectrographs for  imaging  photon spectra from x-ray  sources  in a single shot.
    (a) Spectral filter spectrograph with Bragg reflecting crystal C and focusing element F.  (b) Angular dispersive spectrograph with dispersing element D --- a Bragg reflecting crystal with reflecting atomic planes at a nonzero angle to the crystal surface ---  and focusing element F.}
\label{fig1}
\end{figure}

To determine the performance of a source having such a narrow bandwidth, a new diagnostic tool is required --- a spectrograph capable of imaging photon spectra
in a single measurement. This spectrograph will measure CBXFEL spectra with meV
resolution and high luminosity on a shot-to-shot basis.

Spectrographs have been designed \cite{YHZ06,ZCF12} and demonstrated
\cite{YHZ06,ZCF12,TBKS16,BSR17} and are in use as hard x-ray spectral
diagnostic tools \cite{ZCF12,KFL20} for the current generation of high-gain, single-pass XFELs
\cite{EAA10,HXRSS12,SACLA,IOH19,EXFEL20,NMO21,LGG23}. These devices
use x-ray Bragg diffraction from crystals; namely, they exploit the fact that when diffraction occurs at particular incidence angle $\theta$ (Bragg's
angle), x-rays of a specific energy are filtered out (reflected). 
That is, for photon incidence at Bragg's angle $\theta$ to reflecting atomic planes with interplanar distance
$d_{\ind{H}}$, exclusively x-rays of specific photon energy
$E=E_{\ind{H}}/\sin\theta$ (Bragg's law) are reflected, where
$E_{\ind{H}}=hc/2d_{\ind{H}}$.  Figure~\ref{fig1}(a) shows a
schematic of such a ``spectral filter'' spectrograph.
Despite a very small angular divergence ($\simeq \mu$rad) of the XFEL
beams, these devices can have a significant spectral window of
imaging of $\Delta E_{\ind{\cup}}\simeq E\Omega/\tan\theta \simeq
50$~eV. This large spectral window is a consequence of  artificially introducing a large variation ( $\Omega \simeq
$~1~mrad) in the incidence angle $\theta$ either by focusing x-rays on
the crystal \cite{YHZ06} or by bending the crystal \cite{ZCF12}. The
energy resolution of such spectrographs is limited by the Bragg
reflection bandwidth $\Delta E$, which is typically $\simeq 1-0.1$~eV. The
corresponding small angular acceptance ($\Delta\theta=(\Delta
E/E)\tan\theta\simeq 1-10~\mu$rad) results in low luminosity ($\propto\Delta\theta/\Omega\simeq 10^{-3}-10^{-2}$).

Spectrographs for CBXFELs must have a better spectral resolution
and higher luminosity than available from current spectral filter spectrographs in use at XFELs.  To this end, we apply here an alternative
spectrographic approach that uses angular dispersion of x-rays in Bragg
diffraction from a crystal \cite{Shvydko-SB,SLK06} --- a hard x-ray grating effect --- combined with focusing \cite{KCR09,Shvydko15}. 

The operating principle and basic components of the angular dispersive spectrograph are presented schematically in Fig.~\ref{fig1}(b).  Dispersing element D
is a broadband Bragg crystal reflector in asymmetric scattering
geometry with reflecting atomic planes at nonzero (asymmetry) angle
$\eta$ to the crystal surface. This dispersing element reflects x-rays of different photon
energies $E$ at different angles $\theta^{\prime}$ with the angular
dispersion rate $\dirate={\mathrm d}\theta^{\prime}/{\mathrm d}E$
similar to a diffraction grating (the angular dispersion effect). Finally, 
focusing element F focuses different spectral components on different
locations on the image plane, located at distance $l_{\ind{3}}$ from the
dispersing element. In the general case, the dispersing element can be
composed of several crystals (as in the present paper) exhibiting a
cumulative dispersion rate $\fcomm{}$ larger than that of a single
reflector \cite{SSM13}. Unlike the spectral filtering spectrographs in
Fig.~\ref{fig1}(a), all photons from the source are captured (within the spectral or angular acceptance ranges of the optics), thus
ensuring high luminosity. In addition, the spectral resolution
\begin{equation}
  \Delta \varepsilon = \frac{\Delta y}{|G|}, \hspace{1cm} G = \fcomm{} l_{\ind{3}} 
\label{eq0010}
\end{equation}  
relies on the magnitude of the linear dispersion rate $G={\mathrm
  d}y/{\mathrm d}E$ in the image plane and on the tightness of the
monochromatic focal spot size $\Delta y$ (see \cite{Shvydko15}, Appendix, and
Eq.~\eqref{eq0430}) rather than on the smallness of the Bragg
reflection bandwidth.

A proof-of-principle angular dispersive hard x-ray spectrograph was
demonstrated in \cite{SSM13}, where it was also shown that multiple
crystal arrangements can enhance the cumulative angular dispersion
rate of the system $\fcomm{}$, which is critical for achieving high spectral
resolution (Eq.~\eqref{eq0010}).  Practical spectrographs were
demonstrated for applications in nuclear resonance \cite{CSB18} and
resonant inelastic scattering (RIXS) \cite{BMF21}
spectroscopies.

The spectrograph presented in this publication was designed to
characterize x-ray pulses of a CBXFEL  demonstrator \cite{MAA19,LPM24,LPS24}, which is being developed
through a joint project of Argonne National Laboratory, SLAC
National Accelerator Laboratory, and RIKEN.
The CBXFEL will be driven by electron beams of the Linac Coherent Light Source (LCLS)
facility at SLAC \cite{EAA10} and operated at a fixed nominal photon
energy of $E=$9.831~keV to produce x-ray pulses with a bandwidth of
$\lesssim$~60~meV.  These parameters determine the design values of the spectrograph:
photon energy $E$, spectral window of imaging $\Delta
E_{\ind{\cup}}\gtrsim$~100~meV, and spectral resolution $\Delta
\varepsilon \simeq$~10~meV. In the following discussion, we
present design details of the spectrograph and results of tests performed
at x-ray optics beamline 1-BM at the Advanced Photon Source (APS) at Argonne National Laboratory
\cite{Macrander2016}.

\section{Optical design and components of the spectrograph}

\begin{figure}[t!]
    \includegraphics[width=0.5\textwidth]{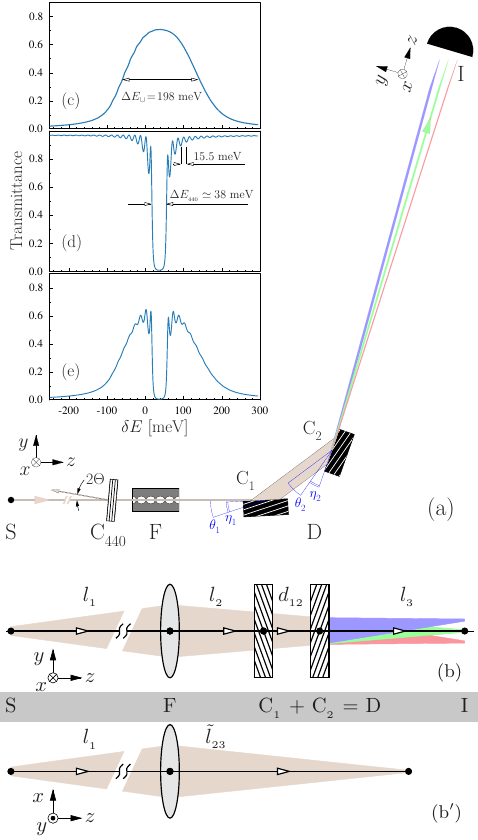}
   \caption{Schematics and spectral profiles of the angular dispersive spectrograph.  (a)
     Optical scheme shown in the \yzplane\ $(y,z)$
      plane with x-ray source S, focusing element F,
double-crystal dispersing element
     D composed of crystals C$_{\ind{1}}$-C$_{\ind{2}}$,
     and x-ray imager I. Crystal C$_{\ind{440}}$ is a spectral resolution probe; see part (d). (b)-(b$^{\prime}$)
     Equivalent unfolded optical schemes in the \yzplane\ (b) and $(x,z)$
     sagittal (b$^{\prime}$) planes. (c)-(e) Spectral profiles of
     x-rays passing through (c) crystals C$_{\ind{1}}$-C$_{\ind{2}}$; (d)
    a spectral resolution probe --- a narrow-band diamond
     Bragg back-reflecting crystal C$_{\ind{440}}$; and (e) through
     both C$_{\ind{440}}$ and C$_{\ind{1}}$-C$_{\ind{2}}$. See text for details.}
\label{fig2}
\end{figure}

\begin{table}
\begin{center}
\resizebox{\columnwidth}{!}{ 
\begin{tabular}{cccc}
     \toprule
     \multicolumn{1}{|c|}{\rule{0pt}{2.6ex}Parameter(s)}  & \multicolumn{1}{|c|}{\rule{0pt}{2.6ex}Notation or} & \multicolumn{2}{|c|}{\rule{0pt}{2.6ex}Value}\\ 

    \cline{3-4} 

    \multicolumn{1}{|c|}{}  & \multicolumn{1}{|c|}{expression} & \multicolumn{1}{|c|}{\rule{0pt}{2.6ex} $1$BM} & \multicolumn{1}{|c|}{\rule{0pt}{2.6ex}LCLS}\\ 
     \toprule
     \toprule
    \multicolumn{4}{|c|}{\textcolor{blue}{X-ray source}}\\
    \toprule
    
    \multicolumn{1}{|c|}{Size (FWHM) [$\mu$m]}  & \multicolumn{1}{|c|}{\rule{0pt}{2.6ex}$\Delta x/\Delta y$} & \multicolumn{1}{|c|}{\rule{0pt}{2.6ex}$198/78$} & \multicolumn{1}{|c|}{\rule{0pt}{2.6ex}$60/60$}\\ 
    
    \cline{3-4}
    \multicolumn{1}{|c|}{Photon energy [keV]}  & \multicolumn{1}{|c|}{$E$} & \multicolumn{2}{|c|}{\rule{0pt}{2.6ex}$9.831$ keV}\\ 

    \toprule
     \multicolumn{4}{|c|}{\textcolor{blue}{\rule{0pt}{2.6ex}Dispersing element D and crystals C$_1$, C$_2$}}\\

     \toprule
    \multicolumn{1}{|c|}{Bragg angles}  & \multicolumn{1}{|c|}{\rule{0pt}{2.6ex}$\theta_{1}=\theta_{2}$} & \multicolumn{2}{|c|}{\rule{0pt}{2.6ex}$18.38^{\circ}$}\\

    \multicolumn{1}{|c|}{Asymmetry angles}  & \multicolumn{1}{|c|}{\rule{0pt}{2.6ex}$\eta_{1}=-\eta_{2}$} & \multicolumn{2}{|c|}{\rule{0pt}{2.6ex} $-16.4^{\circ}$}\\

    \multicolumn{1}{|c|}{Angular acceptance [$\mu$rad]}  & \multicolumn{1}{|c|}{\rule{0pt}{2.6ex}$\Delta\theta_{1}$, $\Delta\theta_{2}$} & \multicolumn{2}{|c|}{\rule{0pt}{2.6ex} 196, 12}\\

    \multicolumn{1}{|c|}{Asymmetry factors}  & \multicolumn{1}{|c|}{\rule{0pt}{2.6ex}$b_{1}$, $b_{2}$  \eqref{fm1006}} & \multicolumn{2}{|c|}{\rule{0pt}{2.6ex}$-0.061$, $-16.5$}\\

    \multicolumn{1}{|c|}{Cumulative factor}  & \multicolumn{1}{|c|}{\rule{0pt}{2.6ex}$\dcomm{}=b_{\ind{1}}b_{\ind{2}}$ \eqref{fm1004}} & \multicolumn{2}{|c|}{\rule{0pt}{2.6ex}$1$}\\

     \multicolumn{1}{|c|}{Angular dispersion}  & \multicolumn{1}{|c|}{\rule{0pt}{2.6ex}$\dirate_{\ind{1}},\dirate_{\ind{2}}$ \eqref{fm1006} } & \multicolumn{2}{|c|}{\rule{0pt}{2.6ex} $-0.032,\, +0.52$}\\

     \multicolumn{1}{|c|}{ rates [$\mu$rad/meV]}  & \multicolumn{1}{|c|}{ } & \multicolumn{2}{|c|}{}\\  

      \multicolumn{1}{|c|}{Cumulative disper-}  & \multicolumn{1}{|c|}{\rule{0pt}{2.6ex}$\fcomm{}=\dirate_{\ind{1}}b_{\ind{2}}+\dirate_{\ind{2}}$  \eqref{fm1004} } & \multicolumn{2}{|c|}{\rule{0pt}{2.6ex} $1.05$}\\

     \multicolumn{1}{|c|}{sion rate [$\mu$rad/meV]}  & \multicolumn{1}{|c|}{ } & \multicolumn{2}{|c|}{}\\  

      \multicolumn{1}{|c|}{Spectral window}  & \multicolumn{1}{|c|}{\rule{0pt}{2.6ex} $\Delta E_{\ind{\cup}}$} & \multicolumn{2}{|c|}{\rule{0pt}{2.6ex} $198$}\\

      \multicolumn{1}{|c|}{of imaging [meV]}  & \multicolumn{1}{|c|}{ } & \multicolumn{2}{|c|}{}\\
      \toprule

      \multicolumn{4}{|c|}{\textcolor{blue}{\rule{0pt}{2.6ex}Focusing element F (CRL) and lenses}}\\

      \toprule
      \multicolumn{1}{|c|}{Be-lens radius [$\mu m$]}  & \multicolumn{1}{|c|}{\rule{0pt}{2.6ex} $R$} & \multicolumn{2}{|c|}{\rule{0pt}{2.6ex} $200$}\\

      \multicolumn{1}{|c|}{Number of lenses}  & \multicolumn{1}{|c|}{\rule{0pt}{2.6ex} $N_{_L}$} & \multicolumn{2}{|c|}{\rule{0pt}{2.6ex} $16$}\\

      \multicolumn{1}{|c|}{Focal length [$m$]}  & \multicolumn{1}{|c|}{\rule{0pt}{2.6ex} $f\!=\!R/(2 N_{\indrm{L}}\delta$\footnote{Refractive index $n=1-\delta$, where decrement $\delta$ = 3.52$\times 10^{-6}$ in Be.}) \cite{LST99} } & \multicolumn{2}{|c|}{\rule{0pt}{2.6ex} $1.772$}\\
      \toprule

      \multicolumn{4}{|c|}{\textcolor{blue}{\rule{0pt}{2.6ex}Spectrograph}}\\
      \toprule

      \multicolumn{1}{|c|}{Distances between}  & \multicolumn{1}{|c|}{\rule{0pt}{2.6ex} $l_{\ind{1}}$} & \multicolumn{1}{|c|}{\rule{0pt}{2.6ex}$\simeq$ $35.1$} & \multicolumn{1}{|c|}{\rule{0pt}{2.6ex}$\simeq$ $250$}\\
      \cline{3-4}
      
      \multicolumn{1}{|c|}{elements [$m$]}& \multicolumn{1}{|c|}{\rule{0pt}{2.6ex} $l_{\ind{3}}$, $d_{\ind{12}}$ } & \multicolumn{2}{|c|}{\rule{0pt}{2.6ex} $1.39$,$\;$  $0.15$}\\
      \cline{3-4}

      \multicolumn{1}{|c|}{}& \multicolumn{1}{|c|}{\rule{0pt}{2.6ex}$\tilde{l}_{\ind{23}}\!=\!fl_{\ind{1}}/(l_{\ind{1}}\!-\!f)$}  & \multicolumn{1}{|c|}{\rule{0pt}{2.6ex}$1.87$}   & \multicolumn{1}{|c|}{\rule{0pt}{2.6ex}$1.78$}  \\ 
      
      \multicolumn{1}{|c|}{}& \multicolumn{1}{|c|}{\rule{0pt}{2.6ex}$l_{\ind{2}}=\tilde{l}_{\ind{23}} - l_{\ind{3}}$ \eqref{fm1031}}  & \multicolumn{1}{|c|}{\rule{0pt}{2.6ex} $0.48$ }  & \multicolumn{1}{|c|}{\rule{0pt}{2.6ex}$0.39$}\\ 

      \multicolumn{1}{|c|}{De-magnification}& \multicolumn{1}{|c|}{\rule{0pt}{2.6ex} $A\!=\!-\tilde{l}_{\ind{23}}/l_{\ind{1}}$ \eqref{fm1012}}  & \multicolumn{1}{|c|}{\rule{0pt}{2.6ex} $0.053$ }  & \multicolumn{1}{|c|}{\rule{0pt}{2.6ex}$0.007$}\\

      \multicolumn{1}{|c|}{Mono. image size [$\mu m$]}& \multicolumn{1}{|c|}{\rule{0pt}{2.6ex}  $\Delta x^{\prime}$/$\Delta y^{\prime}$}  & \multicolumn{1}{|c|}{\rule{0pt}{2.6ex} $10.5/4.1$ }  & \multicolumn{1}{|c|}{\rule{0pt}{2.6ex}$0.4/0.4$}\\
      \cline{3-4}

      \multicolumn{1}{|c|}{Linear dispersion}& \multicolumn{1}{|c|}{\rule{0pt}{2.6ex}  $\gcomm{}=\fcomm{}l_{\ind{3}}$}  & \multicolumn{2}{|c|}{\rule{0pt}{2.6ex} $1.46$ } \\
      
      \multicolumn{1}{|c|}{rate [$\mu m/meV$]} & \multicolumn{1}{|c|}{} &\multicolumn{2}{|c|}{}\\

      \cline{3-4}

      \multicolumn{1}{|c|}{Spectral resolution}& \multicolumn{1}{|c|}{\rule{0pt}{2.6ex}  $\Delta\varepsilon$}  & \multicolumn{2}{|c|}{}\\
      \cline{3-4}

      \multicolumn{1}{|c|}{$-$Ultimate [meV]}& \multicolumn{1}{|c|}{\rule{0pt}{2.6ex} $=\Delta y/(l_{\ind{1}}\fcomm{})$ }  & \multicolumn{1}{|c|}{\rule{0pt}{2.6ex} $2.2$ }  & \multicolumn{1}{|c|}{\rule{0pt}{2.6ex}$0.24$}\\
      \cline{3-4}

      \multicolumn{1}{|c|}{$-$Expected [meV]}& \multicolumn{1}{|c|}{\rule{0pt}{2.6ex}  $=\Delta y_{\indrm {e}}^{\prime}/\gcomm{}$ }  & \multicolumn{2}{|c|}{8.3\footnote{Here, the actual monochromatic focal spot size $\Delta y_{\indrm {e}}^{\prime}$=12~$\mu$m is used to calculate the expected resolution; see Fig.~\ref{fig3}(a).}} \\
      \toprule
           
\end{tabular}
}

\caption{Design parameters of x-ray sources (1-BM at APS or LCLS at
  SLAC); dispersing element D; focusing element F; and spectrograph.}
\label{tab001}
\end{center}
\end{table}

To achieve the needed spectral resolution of $\Delta \varepsilon \simeq 10$~meV,
a dispersing element D with an angular dispersion rate of
$\dirate\simeq 1~\mu$rad/meV is required (see Eq.~\eqref{eq0010}),
assuming $\Delta y \simeq 10~\mu$m and $l_{\ind{3}}\simeq 1$~m.  This performance 
can be realized by using a rather simple dispersing element D composed of
two asymmetrically cut crystals C$_{\ind{1}}$ and C$_{\ind{2}}$ in the
mirror-symmetric dispersive (++) arrangement, as shown in the optical
scheme of the spectrograph in
Fig.~\ref{fig2}(a). The spectrograph is shown in the \yzplane\
plane $(y,z)$, which coincides with the  angular
dispersion plane.

\begin{figure}[t!]
      \includegraphics[width=0.5\textwidth]{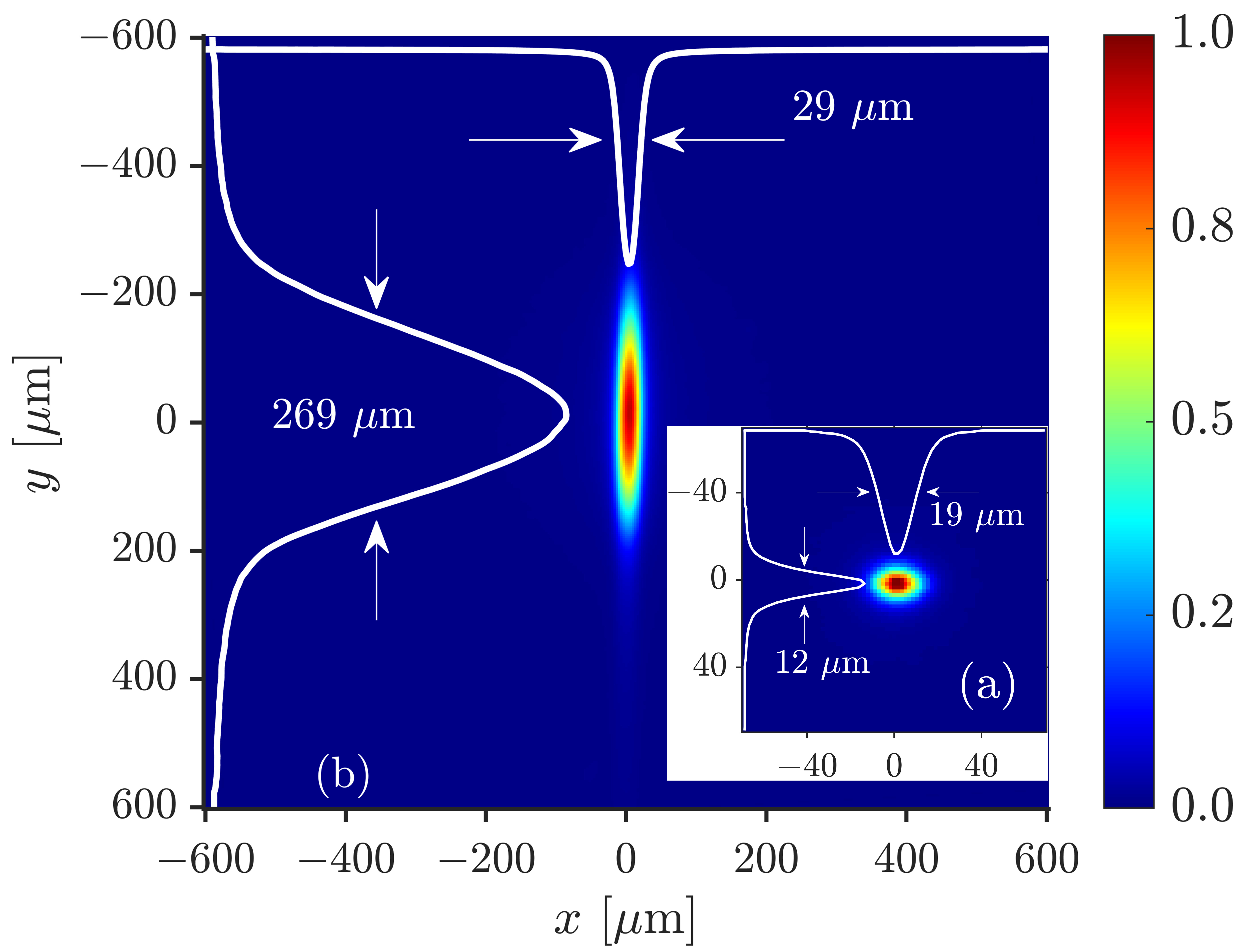}  
\caption{Images of the monochromatized x-ray beam from a bending magnet source take in (a, inset) inline focusing configuration S-F-I  of Fig.~\ref{fig2}(b$^{\prime}$), and   (b) in spectrograph configuration S-F-C$_{\ind{1}}$-C$_{\ind{2}}$-I of Figs.~\ref{fig2}(a)-(b) with  the double-crystal element D (C$_{\ind{1}}$-C$_{\ind{2}}$) dispersing x-rays  in the \yzplane\ plane ($y$-direction). Also shown are beam profiles $S(y)$ and $S(x)$, which are a result of integrating the images over  $x$ and $y$, respectively.}
\label{fig3}
\end{figure}
\begin{figure*}
    \includegraphics[width=0.999\textwidth]{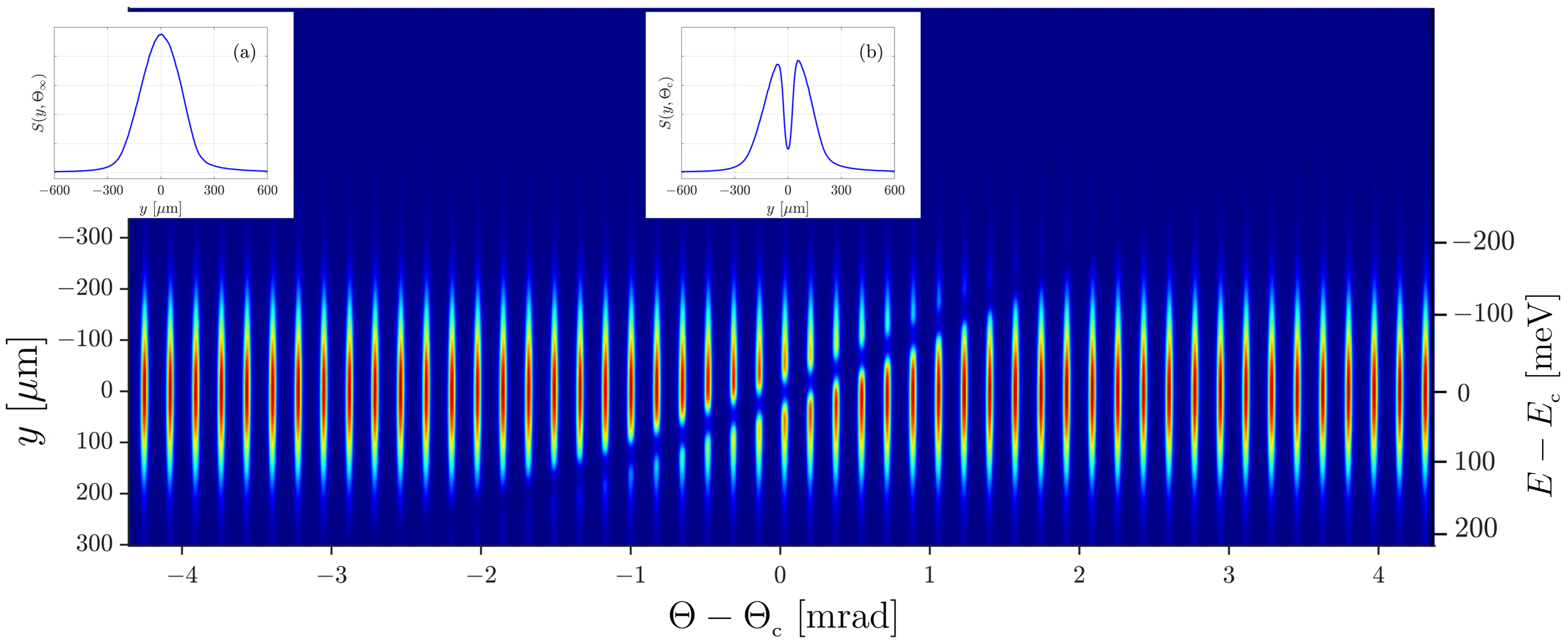}
\caption{A sequence of x-ray spectrograph images similar to that in Fig.~\ref{fig3}(b)  but taken here with
 spectral resolution probe  $C_{\ind{440}}$  in the  beam and  
angular deviation   $\Theta$ from the exact Bragg  back reflection being changed incrementally. Inserts show examples of spectral profiles
$S(y,\Theta)$ taken at selected angles
$\Theta_{\ind{\infty}}=4.91$~mrad (inset a) and
$\Theta_{\indrm{c}}=9.26$~mrad (inset b).
The $\Theta$ scale is centered at $\Theta_{\indrm{c}}=9.259$~mrad; this value corresponds to the notch in the middle of the spectral window of imaging (inset b).  The energy scale $E-E_{\indrm{c}}=  G^{\indrm{(exp)}} y$ is obtained using the linear dispersion rate $G^{\indrm{(exp)}}$ derived from these images  as detailed in the discussion of Fig.~\ref{fig5}. }
\label{fig4}
\end{figure*}

To achieve the target value
of the cumulative dispersion rate $\fcomm{}$, the low-index 220 Bragg reflections from Ge crystals are used, with the
parameters provided in Table~\ref{tab001}.  The use of Ge, rather than the more standard Si, makes it possible to 
maximize the spectral window of imaging to $\Delta E_{\ind{\cup}} =
198$~meV and the angular acceptance. The
calculated spectral profile of the x-rays reflected by the
crystals\footnote{The calculations are performed for x-rays with a
100-$\mu$rad divergence introduced by focusing lens F, which is smaller that the angular acceptance of the dispersing element.}  is shown in
Fig.~\ref{fig2}(c).  The crystals were manufactured at the APS.
Dispersing elements in the same configuration were used previously in
\cite{CSB18,BMF21}. This is the simplest configuration that ensures
enhancement of the cumulative rate $\fcomm{}$ vs. 
dispersion rate  of a single reflector.  Simultaneously, it leaves unchanged the beam cross-section at the exit of the dispersing element, as the cumulative asymmetry factor $\dcomm{}$ is 1 in this
case.

The focusing element F is a compound refractive lens (CRL)
\cite{SKSL} composed of paraboloidal beryllium lenses \cite{LST99}
with parameters presented in Table~\ref{tab001}.  

The spectrograph
design parameters also shown in Table~\ref{tab001}   are optimized using the equations of the spectrograph theory \cite{Shvydko15} that are presented in a focused form in Appendix~A. We also show there that the nonzero distance
$d_{\ind{12}}$ between crystals C$_{\ind{1}}$ and C$_{\ind{2}}$ of the
dispersing element leads to astigmatism: the focal
points of x-rays propagating in the mutually perpendicular $(y,z)$ \yzplane\
and $(x,z)$ sagittal planes are spaced by $\simeq d_{\ind{12}}$ along the
optical axes, as illustrated on the equivalent unfolded
schemes of the spectrograph in Figs.~\ref{fig2}(b) and (b$^{\prime}$).

To characterize the spectrograph's performance --- dispersion rates,
spectral resolution, spectral window of imaging --- a spectral
resolution probe C$_{\ind{440}}$ is added to the setup, as shown in
Fig.~\ref{fig2}(a). Crystal C$_{\ind{440}}$ is an x-ray-transparent diamond crystal of
thickness $d$ = 40~$\mu$m  in the 440 Bragg reflection close to backscattering
with incidence angle $\Theta \ll 1$. Inserting this crystal creates an absorption notch in the x-ray transmission spectrum having a bandwidth $\Delta E_{\ind{440}}$ = 38~meV of
the Bragg reflection. 
The tails of the notch display fringes of equal thickness with a periodicity of $hc/2d$=15.5~meV. The
calculated spectrum is shown in Fig.~\ref{fig2}(d).
Figure~\ref{fig2}(e) presents the result of the combined action of
C$_{\ind{1}}$-C$_{\ind{2}}$ and C$_{\ind{440}}$ on the spectrum of
x-rays propagating through the system. The energy variation of the absorption notch location, $\delta
E/E=\tan\Theta\, \delta \Theta $, is
controlled by variation of the incidence angle $\delta \Theta$
according to Bragg's law; this relationship enables energy calibration of the
spectrograph. The backscattering geometry ensures that the spectral profile of the  notch is insensitive to the angular divergence of the x-ray
beam and also permits straightforward measurement of $\Theta$.

In the experiment, a double-crystal Si(111) monochromator is used 
upstream of C$_{\ind{440}}$ to reduce the bandwidth of x-rays to about 1~eV \cite{Macrander2016}. This monochromator is not shown in Fig.~\ref{fig2}(a). The spectral flux density of photons provided to the experiment is $\simeq 10^9$~ph/s/eV/mm$^2$.

\section{Experiment}
In the first measurement, the monochromatic x-ray source image size was
determined to assess the energy resolution that
can be achieved under the given experimental conditions of beamline 1-BM
 \cite{Macrander2016} at APS.  Figure~\ref{fig3}(a) shows
the source image with a full width at half maximum (FWHM) of $\Delta
x_{\indrm{e}}^{\prime} \times \Delta y_{\indrm{e}}^{\prime} = 19\times
12$~$\mu$m$^2$.  It was measured in the inline S-F-I configuration
excluding the dispersing element as in the equivalent scheme of
Fig.~\ref{fig2}(b$^{\prime}$), with $l_{\ind{1}}$ and
$\tilde{l}_{\ind{23}}$ values from Table~\ref{tab001}.  Given the 1-eV bandwidth of the x-rays,  focusing by the CRL can be considered
achromatic, and therefore the measured sizes  correspond to the
monochromatic source image size.  Both horizontal and vertical values, and especially the
vertical [see Fig. 3(a)], are larger than the design monochromatic image size
$\Delta x^{\prime} \times \Delta y^{\prime}$ given in
Table~\ref{tab001}. This broadening occurs mostly because of imperfections
in the upstream beamline optical components, e.g., the Si(111)
monochromator and Be windows, and the limited spatial resolution $\simeq$~6~$\mu$m  of the  YaG:Ce-based scintillator x-ray imager. The imager uses a low-noise Andor Zyla~4.2~Plus camera and Infinity KC~VideoMax optic with IF‐3.5 objective \cite{LPM24}. The vertical size determines the smallest spectral resolution that can be
expected under conditions of the present experiment: $\Delta \varepsilon_{\indrm{e}} \simeq 8$~meV (see Table~\ref{tab001}).

In the next step, the dispersing element was added to complete the
spectrograph.  Figure~\ref{fig3}(b) demonstrates its immediate
effect: the $y$-image size is greatly enlarged from 12~$\mu$m
to 269~$\mu$m (FWHM). The image was taken in the location of the
\yzplane\ plane focus, which is away from the sagittal focus by about
$d_{\ind{12}}$ = 15~cm due to astigmatism in the system. (The best focusing in the
\yzplane\ and sagittal planes is determined by lens
equations \eqref{fm1033} and \eqref{fm1034}, respectively; see also the numerical simulations discussed in the next section.) Because of this difference in focal position, the $x$-image size
in Fig.~\ref{fig3}(b) is larger than that in Fig.~\ref{fig3}(a).  The
pitch angles of crystals C$_{\ind{1}}$ and C$_{\ind{2}}$
(corresponding to $\theta_{\ind{1}}$ and $\theta_{\ind{2}}$) were
aligned by maximizing the intensity of reflected x-rays, while the yaw and roll angles were aligned to minimize the $x$-image size of the source in the sagittal image plane, which is unaffected by the angular
dispersion effect.
The images were made with $\simeq 10^7$ photons (exposure time 10~s), and a signal-to-noise ratio of $\simeq$~150.

Finally, the spectral resolution probe C$_{\ind{440}}$ was added to the
setup to measure the dispersion rates, the spectral resolution, and
the window of imaging of the spectrograph. Figure~\ref{fig4} shows a
sequence of x-ray spectrograph images similar to that in
Fig.~\ref{fig3}(b) but taken here with the spectral probe in the
beam and the angular deviation $\Theta$ from exact
Bragg back reflection being changed incrementally. As already noted, the effect of C$_{\ind{440}}$ is to produce an
absorption notch. The spectral location of this notch (imager's coordinate $y$)
varies with $\Theta$ according to Bragg's law. The $\Theta$-scale is
centered at $\Theta_{\indrm{c}}=9.259$~mrad; at this angle, the notch is in
the middle of the spectral window of imaging of the
spectrograph. Also shown are examples of spectral image profiles
$S(y,\Theta)$ taken at selected angles
$\Theta_{\ind{\infty}}=4.91$~mrad and
$\Theta_{\indrm{c}}=9.26$~mrad [Figure \ref{fig4}, insets (a) and (b), respectively].

\begin{figure}
       \includegraphics[width=0.49\textwidth]{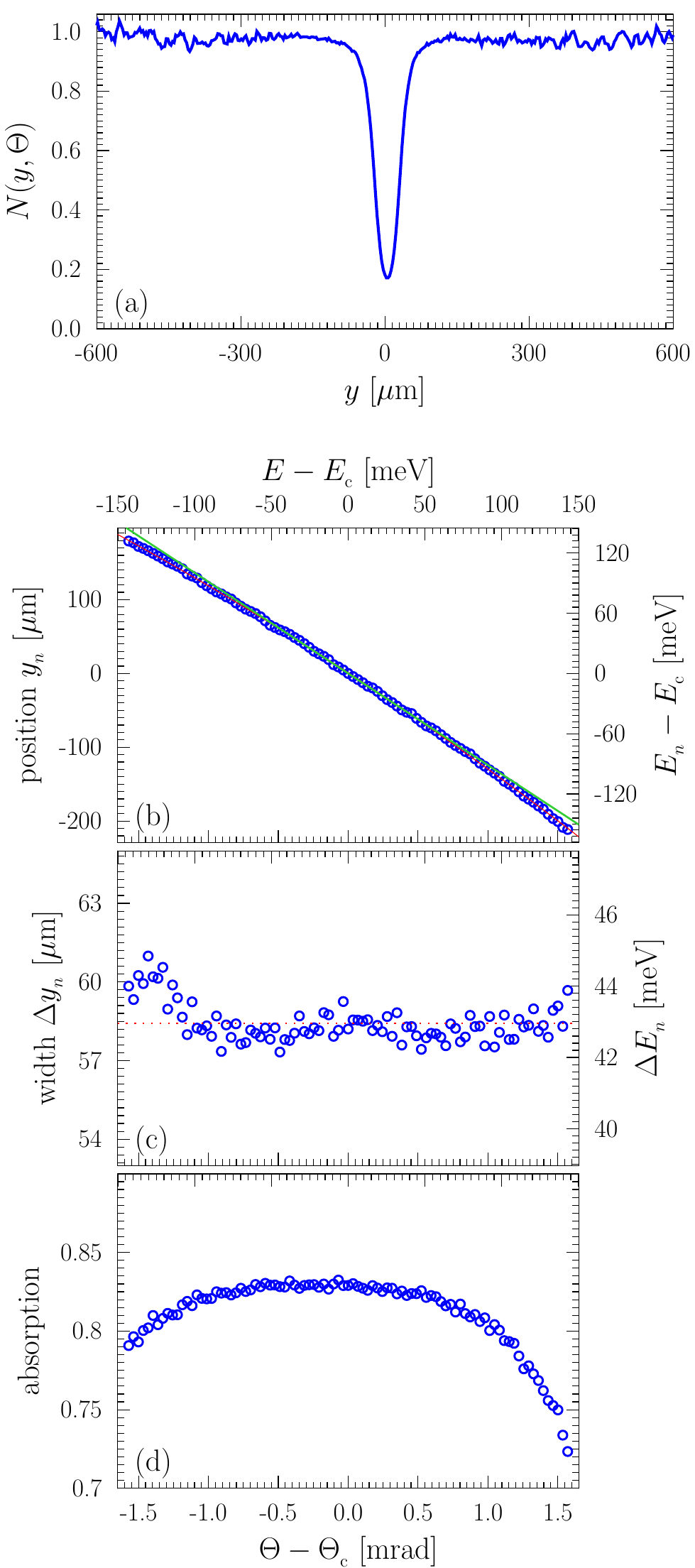}      
\caption{Normalized absorption notch profiles $N(y,\Theta)$ and notch parameters as a function of  angular deviation $\Theta$ from exact Bragg backscattering of probe $C_{\ind{440}}$. (a) An example of the normalized absorption notch profile
  $N(y,\Theta)=S(y,\Theta)/S(y,\Theta_{\ind{\infty}})$: that is, the ratio of the x-ray
  beam spectral image profile $S(y,\Theta)$ [see Fig.~\ref{fig4}(b)] to the
  x-ray beam image profile $S(y,\Theta_{\ind{\infty}})$ when unaffected by the absorption notch [see Fig.~\ref{fig4}(a)].
  (b) Absorption notch location $y_{n}$.
  (c) Notch full width at half maximum (FWHM) $\Delta y_{n}$. (d) Absorption effect.}
\label{fig5}
\end{figure}

  \begin{figure*}[t!]
    \includegraphics[width=0.999\textwidth]{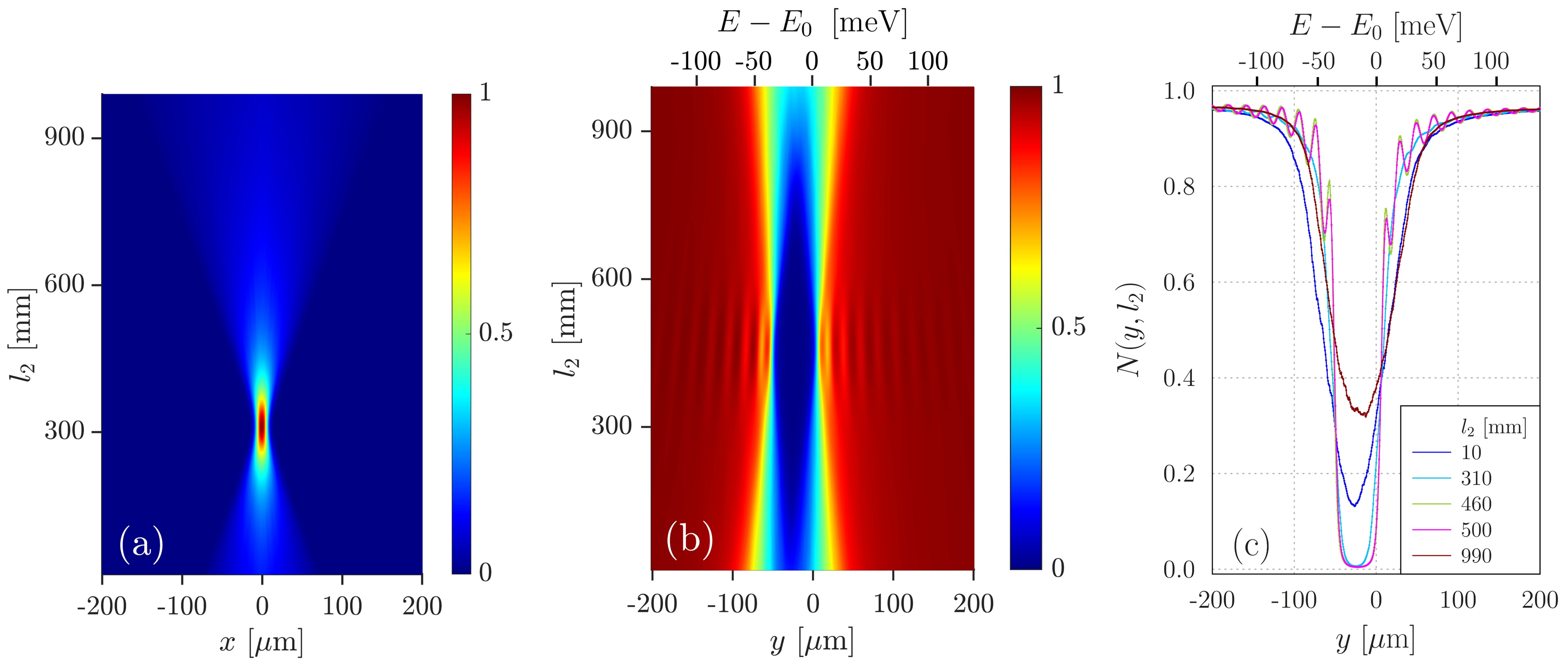}

\caption{Numerical simulations of x-ray imaging by the spectrograph.
  (a) 2D color map of the image profiles $S(x,l_{\ind{2}})$ in the
  sagittal plane as a function of distance $l_{\ind{2}}$ between movable focusing element F (CRL) and fixed crystal C$_{\ind{1}}$. (b) 2D color map of the normalized absorption notch
  profiles $N(y,l_{\ind{2}})$ in the \yzplane\ plane as a function of
  $l_{\ind{2}}$. (c) Examples of the normalized adsorption notch
  profiles $N(y,l_{\ind{2}})$ at particular $l_{\ind{2}}$.} 
\label{fig6}
\end{figure*}

Figure~\ref{fig5} shows an example of a normalized absorption notch
profile $N(y,\Theta)$ in panel (a) and plots of normalized notch parameters in
panels (b)-(d) as a function of angular deviation
$\Theta-\Theta_{\indrm{c}}$ derived from the images of
Fig.~\ref{fig4}. The normalized absorption notch profile
$N(y,\Theta)=S(y,\Theta)/S(y,\Theta_{\ind{\infty}})$ is defined as the ratio of the x-ray beam spectral image
profile $S(y,\Theta)$ at a particular
angle $\Theta$  [see Fig.~\ref{fig4}(b)] to the x-ray beam
image profile $S(y,\Theta_{\ind{\infty}})$ when unaffected by the
absorption notch [see Fig.~\ref{fig4}(a)]. Figures~\ref{fig5}(b)-(d) present, respectively, the  position
$y_{n}$ of the absorption notch minimum, the notch
width $\Delta y_{n}$ (FWHM), and absorption effect, all as function of
$\Theta-\Theta_{\indrm{c}}$, 
evaluated for each normalized absorption notch profile $N(y,\Theta)$.

In
Figs.~\ref{fig5}(b) through \ref{fig5}(d), the bottom angular scale $\Theta-\Theta_{\indrm{c}}$  is converted into the top energy scale
$E-E_{\indrm{c}}$ using Dumond tangent
$\dum=\mathrm{d}E/\mathrm{d}\Theta |_{\indrm{c}} =
E_{\indrm{c}}\tan\Theta_{\indrm{c}}$=91(5)~meV/mrad derived from
Bragg's law at $\Theta=\Theta_{\indrm{c}}$.
The main contribution to the error of $\dum$ is a $\simeq
0.4$~mrad inaccuracy  in determination of $\Theta_{\indrm{c}}$ in the experiment.
The red line
  in Fig.~\ref{fig5}(b) is a quadratic fit.  The green line is its tangent
  $T=\mathrm{d}y_{n}/\mathrm{d}\Theta |_{\indrm{c}}
  =-124~\mu$m/mrad at $\Theta=\Theta_{\indrm{c}}$. The spatial scales
  on the left of Figs.~\ref{fig5}(b) and ~\ref{fig5}(c) and on Fig.~\ref{fig4} are converted to the energy scales on the
  right using the linear dispersion rate of the spectrograph, $G^{\indrm{(exp)}}=
  \mathrm{d}y_{n}/ \mathrm{d}E |_{\indrm{c}} =T/\dum
  =1.36(7)~\mu$m/meV.  The value of the linear dispersion rate $G^{\indrm{(exp)}}$ obtained from the experimental data is in a  good agreement with the design value $G$ of Table~\ref{tab001}.

This result allows us to determine the spectral
window of imaging that corresponds
to the image width $\Delta y$=269~$\mu$m in Fig.~\ref{fig3}(b), namely, $\Delta E_{\ind{\cup}}^{\indrm{(exp)}}$ = 198(10)~meV. This value also agrees well with the design
value $\Delta E_{\ind{\cup}}$ of Table~\ref{tab001}.  Importantly, this result not only present the 
characteristic  spectral range of the x-rays transmitted through the spectrograph but also identify
the region in which the normalized notch profile has a constant width
and relatively  constant absorption effect [see Figs.~\ref{fig5}(c) and \ref{fig5}(d)], that is, the region where 
the spectral imaging has its highest fidelity.

Accordingly, the notch width measured by the imager, 
$\Delta y_{n}=58.5~\mu$m [Fig.~\ref{fig5}(c)] can be
translated to a spectral notch width of $\Delta
E_{n}$=43(2)~meV. The latter is close to, albeit larger than, the
theoretical value of $\Delta E_{\ind{440}}$=38~meV of
Fig.~\ref{fig2}(d). This difference indicates the spectrograph has a limited spectral resolution, which can be estimated as
$\varepsilon^{\indrm{(exp)}}$ = $\sqrt{(\Delta E_{n})^2-(\Delta
  E_{\ind{440}})^2 }$ = 20(4)~meV. This value is more than a factor of two larger than the
expected value of $\varepsilon_{\indrm{e}}$ = 8.3~meV 
(Table~\ref{tab001}). It is also larger than the 15.5-meV period of the
fringes seen in the theoretical spectral profiles in Figs.~\ref{fig2}(d) and \ref{fig2}(e) and explains why these fringes are
not observed in the experiment, e.g., in Fig.~\ref{fig5}(a).
Moreover, it is due to this limited spectral resolution that the measured
absorption effect is about 83\% [Fig.~\ref{fig5}(d)] rather
than the 99\% expected from theory [Fig.~\ref{fig2}(d)].

\section{Numerical simulations and discussion}

Numerical simulation results of x-ray imaging by this angular dispersive spectrograph, 
shown in Fig.~\ref{fig6}, present its ultimate performance
under idealized conditions of the experiment.  The simulations were
carried out with the x-ray optics modeling package Shadow3
\cite{SHADOW3} in the Oasys environment \cite{RS16,RS17} using
the parameters given in Table~\ref{tab001}. The locations
of the x-ray source, crystals C$_{\ind{1}}$ and C$_{\ind{2}}$, and the
image plane are fixed.  The conditions for the best focusing in the
image plane are determined from variation of distance $l_{\ind{2}}$
between the focusing element F (CRL) and  crystal C$_{\ind{1}}$.
The results of the numerical simulations show that the best focusing
for rays propagating in the sagittal and \yzplane\ planes takes
place at different locations of the focusing element.  The best
focusing in the sagittal plane is revealed from the sharpest image
size at $l_{\ind{2}}\simeq$~310~mm on the 2D color map of image
profiles $S(x,l_{\ind{2}})$ in Fig.~\ref{fig6}(a). The best focusing
in the \yzplane\ plane takes place at $l_{\ind{2}}\simeq$480~mm, as
reflected in the largest absorption effect and in the highest
visibility of the fringes of equal thickness on the tails of the normalized absorption
notch profiles $N(y,l_{\ind{2}})$ on the 2D color map of
$N(y,l_{\ind{2}})$ in Fig.~\ref{fig6}(b).  Examples of the
normalized adsorption notch profiles $N(y,l_{\ind{2}})$ at particular
$l_{\ind{2}}$ are shown additionally in Fig.~\ref{fig6}(c).  These results are
in agreement with lens equations \eqref{fm1033} and
\eqref{fm1034} for focusing in the \yzplane\ and sagittal planes,
respectively, and in agreement with the results of the experiment.

The limited spectral resolution observed in the experiment
--- in particular, the inability of the spectrograph to resolve fringes of equal thickness --- is
likely due to the following two reasons.

First, the x-ray trajectory in the $(y,z)$ diffraction plane is
sensitive to pitch angle instabilities of crystals C$_{\ind{1}}$ and
C$_{\ind{2}}$, resulting in  blurring  of the image profiles $S(y)$. As we
show in Appendix~B, an angular variation $\phi$ of the crystal pitch
angle of either of the crystals results in a shift of the spectral
image by $\delta y^{\prime} \simeq l_{\ind{3}} b_{\ind{2}} \phi$, an effect 
aggravated by the large absolute value of $|b_{\ind{2}}|$. For example,
an angular error of $\phi$ = 1$~\mu$rad of one of the crystals, which
could be caused by vibrations, for example, results in a spatial shift of
$\delta y^{\prime}\simeq 22~\mu$m or in an equivalent energy shift
of $\delta y^{\prime}/G \simeq$~16~meV. Such variations can blur sharp
spectral features and degrade the spectral resolution. Vibrations
were minimized during the experiment but were not eliminated
completely.

Second, imperfect roll and yaw angular alignment of crystals
C$_{\ind{1}}$ and C$_{\ind{2}}$ results in mutually nonparallel
diffraction and dispersion planes, which also degrades
the resolution. Minimizing the image size in the sagittal plane by varying the roll and yaw angles, which was used in the experiment, is a
tedious and equivocal procedure. It could be improved in the future experiments by better
pre-alignment of the crystals.

We note that crystal imperfections are unlikely to be the cause of the
degradation in resolution compared to theory, since the x-ray rocking curve imaging
topography \cite{LBH99,SST16,PWH20} of crystals C$_{\ind{1}}$ and
C$_{\ind{2}}$ revealed high crystal quality: almost theoretical values
for the 440 Bragg reflection width and small Bragg plane slope
variations of only $\simeq$0.1~$\mu$rad (rms) over large crystal areas
of 10$\times$10~mm$^2$.

\section{Summary}

A high-luminosity meV-resolution single-shot hard X-ray spectrograph
was designed as a CBXFEL spectral diagnostic tool to image 9.831-keV x-rays in a $\simeq 200$~meV spectral
window with a spectral resolution of a few meV.  The operational principle
of the spectrograph is angular dispersion of x-rays in Bragg
diffraction from crystals.  The spectrograph operates close to design
specification, exhibiting a linear dispersion rate of 1.36~$\mu$m/meV
and a 200-meV window of high-fidelity spectral imaging. The
experimentally obtained spectral resolution of $\simeq 20$~meV
is limited by high sensitivity to crystal angular instabilities at the optics testing bending magnet beamline 1-BM at the Advanced Photon Source, where it
was tested.

\begin{acknowledgments}

Kwang-Je Kim, Marion White, Lahsen Assoufid (Argonne) and Clement Burns
(WMU) are acknowledged for discussions and support.  Xianrong Huang
and Elina Kazman (Argonne) are acknowledged for manufacturing the
germanium crystals for the dispersing element.  Sergey Terentev and
Vladimir Blank (FSBI TISNCM) are acknowledged for manufacturing the
diamond crystal for the spectral resolution probe.  We are grateful to Michael
Wojcik and Brandon Stone (Argonne) for supporting experiments at APS
beamline 1-BM. Luca Rebuffi (Argonne) is acknowledged for helping to set
up x-ray optics modeling with package Shadow3 in the Oasys
environment.  This research used resources of the Advanced Photon
Source, a U.S. Department of Energy (DOE) Office of Science user
facility at Argonne National Laboratory and is based on research
supported by the U.S. DOE Office of Science--Basic Energy Sciences,
under Contract No. DE-AC02-06CH11357.
\end{acknowledgments}

\appendix

\section{Spectrograph theory with an account of spacing between the dispersive crystals}

A theory of the x-ray angular dispersive spectrograph was introduced
in \cite{Shvydko15}.  The ray-transfer matrix technique was used to
derive the spectral resolution and other characteristics of the
spectrograph for diverse configurations of the crystal dispersing
elements and the focusing and collimating optics.  The same technique is applied here to the
particular case of the spectrograph in the 
``focusing-monochromator-I'' (FM1) configuration considered in the present paper, with one focusing element and
the two-crystal dispersing element, as shown in 
the equivalent optical scheme in Fig.~\ref{fig2}(b). The nonzero
distance  between the crystals is accurately taken into
account, resulting in astigmatism in focusing.

\subsection{General equations}

The spectrograph in the FM1 configuration
consists of the focusing element F with a focal length $f$ at a
distance $l_{\ind{1}}$ from the x-ray source S and a dispersing
element with $N$ crystals C$_{\ind{n}}$ ($n=1,2,...,N$) in asymmetric
Bragg diffraction, each characterized by the asymmetry factor
$b_{\ind{n}}$ and angular dispersion rate $\dirate_{\ind{n}}={\mathrm d}\theta_{\ind{n}}^{\prime}/{\mathrm d}E$:
\begin{equation}
  b_{\ind{n}}=-\frac{\sin(\theta_{\ind{n}}+\eta_{\ind{n}})}{\sin(\theta_{\ind{n}}-\eta_{\ind{n}})}, \hspace{0.5cm} \dirate_{\ind{n}} = -\frac{1+b_{\ind{n}}}{E}\tan\theta_{\ind{n}}.
  \label{fm1006}
\end{equation}  
Here, $\theta$ and $\theta^{\prime}$ are the glancing angle of
incidence (Bragg's angle) and reflection, respectively, to the
diffracting atomic planes; $\eta $ is the asymmetry angle --- the angle
between the diffracting atomic planes and the entrance crystal
surface; $E$ is the x-ray photon energy.  Figure~\ref{fig2}(a) shows
an example of a spectrograph with a two-crystal dispersing element,
while Figs.~\ref{fig2}(b) and \ref{fig2}(b$^{\prime}$) show equivalent unfolded schemes.

In the following, cumulative values are also used for the asymmetry
parameter $\dcomm{n}$ and the angular dispersion rate $\fcomm{n}$; these values 
result from successive Bragg reflection from $n$ crystals ($1\leq n\leq N$), which
are defined as
\begin{equation}
\dcomm{n}=b_{\ind{1}}b_{\ind{2}}b_{\ind{3}} \dotsc b_{\ind{n}}, \hspace{0.5cm}
\fcomm{n}=b_{\ind{n}}\fcomm{n-1} + \sgn_{\ind{n}}\dirate_{\ind{n}}.
\label{fm1002}
\end{equation}
Here $\sgn = -1$ for clockwise and  $\sgn = +1$ for counterclockwise ray deflection upon Bragg reflection.

The first crystal C$_{\ind{1}}$ is at a distance $l_{\ind{2}}$ from
the focusing element F, and the last crystal C$_{\ind{N}}$ is at a
distance $l_{\ind{3}}$ from the image plane with position-sensitive
imager I.  There are nonzero distances $d_{\ind{n-1~ n}}$ between
crystals C$_{\ind{n-1}}$-C$_{\ind{n}}$ of the dispersing element.

Following \cite{Shvydko15}, x-rays of each monochromatic spectral
component $E$ emerging from the source S with a lateral size $\Delta
x \times \Delta y$ are focused to $\Delta x^{\prime}$ in the sagittal $(x,z)$
plane and to $\Delta y^{\prime}$ in the diffraction $(y,z)$ plane:
\begin{equation}
\Delta u^{\prime}\, =\, A\, \Delta u, \hspace{0.15cm} u=x\vee y, \hspace{0.35cm} A = \frac{1}{\dcomm{N}}\left(1- \frac{\tilde{l}_{\ind{23}}}{f} \right), 
\label{fm1010}
\end{equation}
\begin{equation}
\tilde{l}_{\ind{23}} = l_{\ind{23}} + \dcomm{N} \bcomm{N},\hspace{0.5cm} l_{\ind{23}} = l_{\ind{2}} + l_{\ind{3}} \dcomm{N}^2,
\label{fm1020}
\end{equation}
\begin{equation}  
  \bcomm{n}\!=\!\frac{\bcomm{n-1}+\dcomm{n-1}d_{\ind{n-1~ n}}}{b_{\ind{n}}}, \hspace{0.5cm} \bcomm{1}\!=\!0,
\label{fm1023}
\end{equation}  
where, the relationship between $f$, $l_{\ind{1}}$, $l_{\ind{2}}$,
$d_{\ind{n-1~ n}}$, and $l_{\ind{3}}$ is determined by the lens equation for the condition of best focusing: 
\begin{equation}
\frac{1}{l_{\ind{1}}}+\frac{1}{\tilde{l}_{\ind{23}}}=\frac{1}{f}.
\label{fm1030}
\end{equation}
With this equation the magnification factor in Eq.~\eqref{fm1010} can be presented as 
\begin{equation}
\Delta u^{\prime}\, =\, A\, \Delta u, \hspace{0.15cm} u=x\vee y, \hspace{0.35cm} A = -\frac{1}{\dcomm{N}}\,\frac{\tilde{l}_{\ind{23}}}{l_{\ind{1}}}. 
\label{fm1011}
\end{equation}
Importantly, according to Eq.~\eqref{fm1030} the virtual
lens-to-image-plane distance $\tilde{l}_{\ind{23}}$ is solely
determined by $l_{\ind{1}}$, and $f$ and is independent of the details
of the design of the dispersing element.

We assume that Bragg diffraction takes place only in the diffraction $(y,z)$ plane.
However, when considering focusing of x-rays in the sagittal $(x,z)$ plane, which are unaffected by
diffraction in the plane perpendicular to it, we have to use in this
case $b_{\ind{n}}=-1$ and $\dirate_{\ind{n}}=0$ for all crystals.  As
a result, although lens equation \eqref{fm1030} and the virtual 
distance $\tilde{l}_{\ind{23}}$ for the best focusing are the same in
both planes, assuming the same $l_{\ind{1}}$ and $f$, the magnification
factor [Eq.~\eqref{fm1011}] and the image plane locations for x-rays
propagating in the \yzplane\ $(y,z)$ and the sagittal
$(x,z)$ planes [determined by $l_{\ind{2}}$, $d_{\ind{n-1~ n}}$,
$l_{\ind{3}}$ from Eqs.~\eqref{fm1020}-\eqref{fm1023}] can differ.
Therefore, generally speaking, focusing systems that contain dispersing elements exhibit
astigmatism: \yzplane\ and sagittal rays form foci
at different distances along the optic axis.

A change $\delta E$ in the x-ray photon energy $E$ results in a change of the focal spot location in the diffraction $(y,z)$ plane by 
\begin{equation}
\delta y = G\, \delta E, \hspace{0.5cm} G= \fcomm{N} l_{\ind{3}} + \gcomm{N}, 
\label{fm10402}
\end{equation}
  \begin{equation}  
\gcomm{n}\!=\! \frac{\gcomm{n-1}+\fcomm{n-1}d_{\ind{n-1~ n}}}{b_{\ind{n}}}, \hspace{0.5cm} \gcomm{1}\!=\!0,
\label{fm10422}
  \end{equation}
as a result of the angular dispersion in Bragg diffraction from the
asymmetrically cut crystals. Here, $G$ is the linear dispersion rate of
the spectrograph.

An energy variation, which results in a change $|\delta y|$ of the
location of the source image [Eq.~\eqref{fm10402}]  that is equal to
the monochromatic source image size $\Delta y^{\prime}$ [Eqs.~\eqref{fm1010}-\eqref{fm1030}] determines the  energy
resolution  of the spectrograph:
\begin{equation}
\Delta \varepsilon_{\indrm{u}} =  \frac{\Delta y^{\prime}}{|G|}.
\label{eq0430}
\end{equation} 
Using Eq.~\eqref{fm1011} for the magnification factor, the expression for the 
ultimate spectral resolution of the spectrograph (in
the FM1 configuration) can be presented via the source $y$-size as follows:
\begin{equation}
\Delta \varepsilon_{\indrm{u}} =  \frac{|\Delta y|}{|G \dcomm{N}|} \frac{\tilde{l}_{\ind{23}}}{ l_{\ind{1}}}. 
\label{eq0400}
\end{equation}
The monochromatic source image size $\Delta
y^{\prime}_{\indrm{e}}$ in an actual experiment can be larger than the ultimate value $\Delta
y^{\prime}$ given by Eq.~\eqref{fm1011} because of imperfections in
the optical elements of the spectrograph or limited imager spatial resolution. In this case, the
energy resolution of the spectrograph is given by
\begin{equation}
\Delta \varepsilon_{\indrm{e}} =  \frac{\Delta y^{\prime}_{\indrm{e}}}{|G|}. 
\label{eq0410}
\end{equation}

\subsection{Two-crystal mirror-symmetric arrangement}

In a particular case of the
two-crystal ($N=2$) dispersing element in mirror-symmetric dispersive (++) arrangement schematically presented in
Fig.~\ref{fig2}, the crystal angles are
\begin{equation}
\theta_{\ind{1}}=\theta_{\ind{2}}, \hspace{0.5cm} \eta_{\ind{1}}=-\eta_{\ind{2}}<0.     
\label{fm10040}  
\end{equation}
In this case, using Eq.~\eqref{fm1002},  the cumulative values are given by  
\begin{equation}
\dcomm{2}=b_{\ind{1}}b_{\ind{2}}=1, \hspace{0.5cm}
\fcomm{2}=b_{\ind{2}}\dirate_{\ind{1}} + \dirate_{\ind{2}}.
\label{fm1004}  
\end{equation}
Further, the expression for the virtual lens--to--image plane distance 
\begin{equation}
  \tilde{l}_{\ind{23}} = l_{\ind{2}} + l_{\ind{3}}  + b_{\ind{1}}^2 d_{\ind{1 2}}
\label{eq050}
\end{equation}    
is obtained from Eqs.~\eqref{fm1020}-\eqref{fm1023}  and  $\bcomm{2}={b_{\ind{1}}d_{\ind{12}}}/{b_{\ind{2}}}$.
Furthermore, the linear dispersion rate 
\begin{equation}
G= \fcomm{{2}} l_{\ind{3}}   + \frac{\dirate_{\ind{1}} d_{\ind{1 2}}}{b_{\ind{2}}}.
\label{fm1042}
\end{equation}
is obtained from Eqs.~\eqref{fm10402}-\eqref{fm10422}
and
$\gcomm{2}={\dirate_{\ind{1}} d_{\ind{1
      2}}}/{b_{\ind{2}}}$, while
the expression for the magnification factor  
\begin{equation}
A = -\frac{\tilde{l}_{\ind{23}}}{l_{\ind{1}}}
\label{fm1012}
\end{equation}
is derived from Eqs.~\eqref{fm1011} and \eqref{fm1004}.

If the crystal asymmetry is large, i.e.,
$-b_{\ind{1}}=-1/b_{\ind{2}}\ll 1$, and the distance between the
crystals is small, i.e., $d_{\ind{1 2}} \ll l_{\ind{2}}+l_{\ind{3}}$ (as in
our experiment), then the term $b_{\ind{1}}^2 d_{\ind{1 2}}$ can be
omitted in Eq.~\eqref{eq050}. The linear dispersion
term ${\dirate_{\ind{1}} d_{\ind{1 2}}}/{b_{\ind{2}}}$ in
Eq.~\eqref{fm1042} can be also neglected. In this case, $d_{\ind{1 2}}$
does not enter the virtual lens--to--image plane distance
\begin{equation}
  \tilde{l}_{\ind{23}} \simeq l_{\ind{2}} + l_{\ind{3}},
\label{fm1031}
\end{equation}
and it appears as if distance $d_{\ind{1 2}}$ has no impact either on the lens equation, which now reads  as
\begin{equation}
\frac{1}{l_{\ind{1}}}+\frac{1}{l_{\ind{2}} + l_{\ind{3}}} =\frac{1}{f},
\label{fm1033}
\end{equation}
or on the spectral resolution 
\begin{equation}
\Delta \varepsilon =  \frac{\Delta y\,}{\fcomm{n} l_{\ind{3}}} \frac{l_{\ind{2}} + l_{\ind{3}}}{ l_{\ind{1}}}. 
\label{eq043}
\end{equation}

On the other hand, in the symmetric case when  $b_{\ind{1}}=b_{\ind{2}}=-1$, the virtual lens--to--image plane distance becomes
\begin{equation}
\tilde{l}_{\ind{23}} = l_{\ind{2}} + l_{\ind{3}}  + d_{\ind{12}},
\label{fm1035}
\end{equation}
i.e.,  is exactly equal to the length of the optical path from the lens to the imager, with the lens equation taking the expected form
\begin{equation}
\frac{1}{l_{\ind{1}}}+\frac{1}{l_{\ind{2}} + l_{\ind{3}}  + d_{\ind{12}}}  =\frac{1}{f}.
\label{fm1034}
\end{equation}
Here we emphasize that the virtual distance $\tilde{l}_{\ind{23}}$ is
an invariant. What changes is the relationship between $l_{\ind{2}}$ and
$l_{\ind{3}}$ and thus the location of the image planes. In relation
to our experiment, these results mean that there are two different
image planes for focusing of the rays in the \yzplane\
plane $(y,z)$ and in the sagittal 
$(x,z)$ plane perpendicular to it. The sagittal image plane location is defined by focusing
equation \eqref{fm1034}, while the \yzplane\ image plane location is
defined by focusing equation \eqref{fm1033} and is  $d_{\ind{1
    2}}$ further away from the second crystal than the sagittal image
plane, as shown in the equivalent unfolded schemes in
Figs.~(b) and \ref{fig2}(b$^{\prime}$).

\section{Ray trajectory distortions caused by angular errors of C$_{\ind{1}}$  and C$_{\ind{2}}$ crystals}

In this section we study how much an angular misalignment $\phi$ of
either crystal C$_{\ind{1}}$ or C$_{\ind{2}}$ of the dispersing
element of the spectrograph can change the x-ray trajectory and
therefore the image location in the imager plane. In practical terms,
this knowledge determines  tolerances on angular stability of the
dispersing element crystals.

\subsection{General equation}
The ray-transfer matrix approach can be used to propagate rays through
optical systems with misaligned optical elements \cite{Siegman,QS22}.

\begin{figure}
\includegraphics[width=0.325\textwidth]{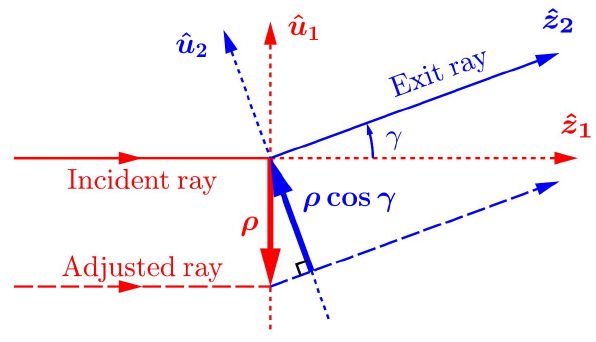}
  \caption{Schematic of incident and adjusted rays clarifying calculation of the x-ray trajectory distorted by a misaligned optical element \cite{QS22}.}
  \label{fig7}
\end{figure}

Figure\(~\)\ref{fig7} (adapted from \cite{QS22})
illustrates the equations used to calculate misalignments. Suppose an
optical element (a crystal or a lens) is displaced from the perfect
configuration by a distance $\rho$ and an angle $\phi$. The distorted
propagation of the x-rays can be calculated by (i) displacing the
incident x-ray presented by vector $\vc{r}_{\ind{1}}$ with spatial and angular coordinates $u_{\ind{1}}$ and $v_{\ind{1}}$, respectively,  by the same amount as the optical element but in the
opposite direction ($-\rho$ and $-\phi$); (ii) transferring the ray
with an appropriate matrix $\hat{M}$ of the optical element; and (iii) moving the transferred
ray back by the adjusted amount. As a result, the 
distorted ray presented by vector $\vc{r}_{\ind{2}}$ with coordinates $u_{\ind{2}}$ and $v_{\ind{2}}$ are given by \cite{QS22}
\begin{align}
  \label{eq010}
  \vc{r}_{\ind{2}}\!=\!
    \begin{bmatrix}
      u_{\ind{2}} \\
      v_{\ind{2}} 
    \end{bmatrix}
    = \hat{M}
    \left(
      \begin{bmatrix}
      u_{\ind{1}} \\
      v_{\ind{1}} 
      \end{bmatrix}
      -
      \begin{bmatrix}
        \rho\\
        \phi
      \end{bmatrix}
    \right)
    +
    \begin{bmatrix}
      \rho\cos\gamma\\
      \phi
    \end{bmatrix} ,
\end{align} 
where $\gamma$ is the angle between the exit ray and the incident ray
at the location of the optical element, as shown in
Fig.\(~\)\ref{fig7}.  For a crystal, the exit ray angle
\(\gamma=2\theta\) after crystal reflection with Bragg's angle $\theta$.
For a lens, $\gamma=0$. Note that the first adjustment of
the ray position $-\rho$ is relative to the incident ray.  To
restore the exit ray to its correct position, it needs to be moved
by the amount $\rho\cos\gamma$.

\subsection{Application to the spectrograph case with two-crystal dispersing element}

We consider a simplified case, in which only a nonzero  angular error $\phi$ is assumed, while the spatial error $\rho=0$. Besides, we will trace  the trajectory of only  the principle ray $\vc{r}_{\ind{1}}\!=\!0$, which is initially on the optical axis. In this case Eq.~\eqref{eq010} simplifies to 
\begin{equation}
\vc{r}_{\ind{2}}\!=\! \left(1-\hat{M} \right) \!\left[
\begin{array}{c} 0 \\ \phi
\end{array}
\right]
\label{eq0220}
\end{equation}

We recall that ray-transfer matrix of an asymmetrically cut crystal is \cite{MK80-2,Shvydko15} 
\begin{gather}
  \hat{C}\,=\,\left[ \begin{array}{cc} 1/b & 0 \\ 0 & b  \end{array} \right], 
\label{eq0122}
\end{gather}
where $b$ is the asymmetry parameter.

In particular, if crystal C$_{\ind{2}}$ is misaligned with an angular error $\phi$, the Bragg reflected ray is
\begin{equation}
  \begin{split}
  \vc{r}_{\ind{2}} & \!=  \! \left(1-\hat{C}_{\ind{2}} \right)  \!\left[ \begin{array}{c} 0 \\ \phi \end{array} \right] \!=\!
\left[ \begin{array}{cc} 1-1/b_{\ind{2}} & 0 \\ 0 & 1-b_{\ind{2}}  \end{array} \right] 
\left[ \begin{array}{c} 0 \\ \phi \end{array} \right] \\
 & \!=\! \left[ \begin{array}{c} 0 \\ (1-b_{\ind{2}}) \phi \end{array} \right].
 \end{split}
\label{eq030}
\end{equation}
This means that the reflected ray propagates at a deflection angle $\Psi_{\ind{2}}=(1-b_{\ind{2}}) \phi$ to the optical axis.
In the symmetric case, for which $b_{\ind{2}}=-1$, the  deflection angle is $\Psi_{\ind{2}}=2\phi$. However,
if $-b_{\ind{2}} \gg 1$ (as in our experiment)  the asymmetrically cut crystal amplifies the deflection angle by a factor $1-b_{\ind{2}} \gg 1$. At a distance $l_{\ind{3}}$, the spatial shift from the optical axis is $l_{\ind{3}}\Psi_{\ind{2}}=l_{\ind{3}}(1-b_{\ind{2}}) \phi$.

If crystal C$_{\ind{1}}$ is misaligned with an angular error $\phi$, the ray after two Bragg reflections (neglecting the spacing $d_{12}$) is
\begin{equation}
  \begin{split}
  \vc{r}_{\ind{2}} & \!=  \! \hat{C}_{\ind{2}} \left(1-\hat{C}_{\ind{1}} \right)  \!\left[ \begin{array}{c} 0 \\ \phi \end{array} \right] \!=\!
\left[ \begin{array}{cc} 1/b_{\ind{2}} & 0 \\ 0 & b_{\ind{2}}  \end{array} \right] 
\left[ \begin{array}{c} 0 \\ (1-b_{\ind{1}}) \phi \end{array} \right] \\
 & \!=\! \left[ \begin{array}{c} 0 \\ b_{\ind{2}} (1-b_{\ind{1}}) \phi \end{array} \right].
 \end{split}
\label{eq0303}
\end{equation}
Although $-b_{\ind{1}}\ll 1 $, and therefore the angular deflection after the first crystal $(1-b_{\ind{1}}) \phi $ is relatively small, the second crystal still amplifies this deflection to $\Psi_{\ind{1}}=b_{\ind{2}} (1-b_{\ind{1}})\phi $. If $-b_{\ind{2}}\gg 1$,  the magnitude of $\Psi_{\ind{1}}$ is close to $\Psi_{\ind{2}}$, but has the opposite sign. Similarly,  the spatial shift from the optical axis at a distance $l_{\ind{3}}$ is $l_{\ind{3}}\Psi_{\ind{1}}=l_{\ind{3}}(1-b_{\ind{1}})b_{\ind{2}} \phi$.

The analytical calculations provided here are confirmed by numerical
simulations with the x-ray optics modeling package Shadow3
\cite{SHADOW3}.

\end{document}